\documentstyle[11pt]{article}\textheight 230mm\textwidth 150mm
            \newcommand{\be}{\begin{eqnarray}}
            \newcommand{\ee}{\end{eqnarray}}
            \newcommand{\eel}[1]{\label{#1}\end{eqnarray}}
\newcommand{\e}[1]{\label{e:#1}\end{eqnarray}}
     \newcommand{\eg}{{\em e.g.\ }}
            \newcommand{\ie}{{\em i.e.\ }}
            \newcommand{\ga}{{\gamma}}

            \newcommand{\del}{{\delta}}
   \newcommand{\dq}{\dot{q}}
   \newcommand{\dt}{\dot{t}}

            \newcommand{\pet}{{\cal P}}

\newcommand{\ca}{{\cal C}}
\newcommand{\baca}{\bar{\cal C}}
            
            \newcommand{\beq}{\begin{quote}}
            \newcommand{\eq}{\end{quote}}
            
            \newcommand{\al}{\alpha}
            \newcommand{\ben}{\begin{enumerate}}
            \newcommand{\een}{\end{enumerate}}
            \newcommand{\bit}{\begin{itemize}}
            \newcommand{\ei}{\end{itemize}}
    	\newcommand{\nn}{\nonumber}
            \newcommand{\r}[1]{(\ref{e:#1})}
            \newcommand{\edfl}[1]{\label{#1}\end{df}}

\newcommand{\vb}{{\cal h}}
\newcommand{\hb}{{\cal i}}

\newcommand{\bapet}{\bar{\pet}}

\newcommand{\bett}{{\bf 1}}

\begin{document}
\begin{titlepage}
\noindent
G\"{o}teborg ITP 96-12\\
August 1996\\
hep-th/yymmddd\\

\vspace*{5 mm}
\vspace*{35mm}
\begin{center}{\LARGE\bf Time evolution in general gauge theories\\ on inner product
spaces}\end{center} \vspace*{3 mm} \begin{center} \vspace*{3 mm}

\begin{center}Robert
Marnelius\footnote{E-mail: tferm@fy.chalmers.se} \\ \vspace*{7 mm} {\sl Institute of
Theoretical Physics\\ Chalmers University of Technology\\
G\"{o}teborg University\\
S-412 96  G\"{o}teborg, Sweden}\end{center}
\vspace*{25 mm}
\begin{abstract}
As previously  shown  BRST singlets $|s\hb$ in a BRST quantization of
general gauge theories on  inner product spaces may be represented in the form
$$|s\hb=e^{[Q, \psi]} |\phi\hb$$ where  $|\phi\hb$   is either a trivially BRST
invariant state  which  only depends on the matter  variables, or
 a solution of a Dirac quantization.
$\psi$ is a corresponding fermionic gauge
fixing operator.  In this paper it is shown that the time evolution
is determined  by the singlet states of
the corresponding reparametrization
invariant theory. The general case when the
constraints and Hamiltonians may have
explicit time dependence is treated.\end{abstract}\end{center}\end{titlepage}

\setcounter{page}{1}
\setcounter{equation}{0}
\section{Introduction.}
In  BRST quantization of general gauge
theories  on inner product spaces the
BRST singlets, $|s\hb$, are BRST invariant states that describe the true physical
degrees of freedom and represent the BRST cohomology ($|s\hb\in \mbox{Ker} Q/\mbox{Im}
Q$) \cite{KO}. By means of a generalized quartet mechanism a simple general
representation of the BRST singlets for general gauge theories with finite number of
degrees of freedom was obtained in  \cite{Solv}. (For Lie group theories corresponding
gauge invariant states were previously obtained by means of a bigrading in
\cite{Simple,Gauge}.) The representation is \be &&|s\hb=e^{[Q,
\psi]}|\phi\hb 
\e{1}
where $Q$ is the hermitian nilpotent BRST  operator in BFV form \cite{BFV,BF1}, 
$\psi$  a  hermitian fermionic gauge fixing operator, and $|\phi\hb$   BRST
invariant states determined by a {\em hermitian} set of operators which are BRST 
doublets in involution. $|\phi\hb$ does not belong to an inner product space although
$|s\hb$ does. Since the BRST quartets of operators may always be split into two sets of
hermitian BRST doublets there are at least two dual choices for $|\phi\hb$ and the
corresponding $\psi$. For general, both irreducible and reducible, gauge theories of
arbitrary rank within the BFV formulation it was found in \cite{Solv} that there
always exist solutions such that  
 $|\phi\hb$
are trivial BRST invariant states  which only depend on the matter 
variables for one set of BRST doublets. For the complementary set of BRST
doublets in the BRST quartets  
 $|\phi\hb$ must be solutions of a Dirac quantization (which not always exist). At the
end of \cite{Solv}  some aspects that needed further clarifications was pointed out
like \eg  the exact connection  to the coBRST formulation and the freedom in the
choice of gauge fixing fermions, as well as the question of how the time evolution in
terms of  a nontrivial Hamiltonian should be defined. The first aspects were clarified
in ref.\cite{CoB}. The second aspect is the subject of the present paper. 

As in \cite{Path} it is natural to expect the time evolution to be given by
\be
&&|s,t\hb=e^{itH_0}|s\hb
\e{2}
in the case when  $H_0$ is a BRST invariant Hamiltonian ($[Q, H_0]=0$) with no explicit
time dependence. On the other hand one could equally well have
\be
&&|\phi, t\hb=e^{itH_0}|\phi\hb,\;\;\;|s,t\hb=e^{[Q,
\psi]}|\phi,t\hb
\e{3}
which is not the same as \r{2} if $H_0$ does not commute with $[Q, \psi]$. 
In \cite{Path}
eq.\r{2} was assumed together with
\be
&&[H_0, [Q, \psi]]=0
\e{4}
which makes \r{2} and \r{3} equivalent. In this case one could bring the
representation \r{1} in contact with the standard path integral formulation in phase
space as formulated in \cite{BFV}.

In this paper we shall determine the time evolution of states of the form \r{1} for
the general case when the Hamiltonian and the gauge fixing fermions $\psi$ in \r{1}
also may have explicit time dependence.
This we shall do without invoking any new ingredients of the BRST quantization as
formulated in \cite{Solv}. What we shall do is to make use of a well-known trick how
to make any theory reparametrization invariant. The resulting reparametrization
invariant theory has then no non-trivial Hamiltonian. Instead it has a new constraint
which involves the original Hamiltonian. The method of \cite{Solv} may therefore be
directly applied to this extended formulation and at the end we may interpret the
resulting BRST singlets  as the time dependent singlets of the original theory by means
of an appropriate gauge choice. In section 2 we define what we mean by the
corresponding reparametrization invariant theory and determines its BRST charge
operator. In section 3 we apply the rules of \cite{Solv} to this extended formulation
and discuss how the gauge fixing should be performed in order for the result to be
interpretable as the time evolution of the original theory. In section 4 we describe
how the formulation looks in the corresponding path integral formulation. In section
5 the simple example of a general regular theory is treated and in section 6 we give
the conclusion.

\setcounter{equation}{0}
\section{Extended formulation}
A general gauge theory with finite number of degrees of freedom may be given in terms
of the phase space Lagrangian ($i=1,\ldots,n;\;\al=1,\ldots,m<n$)
\be
&&L_0(t)=p_i\dq^i-H-v^{\al}\Phi_\al
\e{201}
where $v^\al$ are Lagrange multipliers and $\Phi_\al$ first class constraints. In
distinction to the treatment in ref.\cite{Solv}   $H$
and $\Phi_\al$ are here allowed to depend on time $t$ explicitly. (Such gauge theories
have been  treated in  \cite{BLy,GTy}.) We shall now make use of the fact that the
gauge theory \r{201} may be embedded in a larger gauge theory described by the
Lagrangian \be &&L(\tau)=p_i\dq^i+\pi\dt-v(\pi+H+v^{\al}\Phi_\al)
\e{202}
where we have promoted time $t$ to a dynamical variable $t(\tau)$ with $\pi$ as its
conjugate momentum. (In eq.\r{202} dots  represent differentiation with respect to
the parameter $\tau$!) This is a
well known trick how to make a theory reparametrization invariant (see \eg the books
\cite{GTy,Lan}).  Equivalence between \r{201} and \r{202} requires
$v=dt/d\tau$. We shall require $t$ and $\tau$ to be in a one-to-one correspondence
which demands the condition $v>0$ or $v<0$ on the Lagrange multiplier $v$. (A natural
way to satisfy this condition is to replace $v$ by $e^\omega$ where $\omega$ is
unrestricted.) The Lagrangian \r{202} gives rise to the constraints 
\be 
&&\pi+H=0,\;\;\;\Phi_\al=0
\e{203}
which we require to be of first class so that $v$ and $v^\al$ in \r{202} are
independent variables. Since  $H$ and $\Phi_\al$ do not depend on
$\pi$, the Poisson algebra of the constraints \r{203} becomes
\be
&&\{\Phi_\al, \Phi_\beta\}=c_{\al\beta}^{\;\;\;\ga}\Phi_\ga,\;\;\;\{\Phi_\al,
\pi+H\}=c_\al^{\;\;\beta}\Phi_\beta
\e{204}
where the last condition corresponds to Dirac's consistency conditions for the
original theory.

 We turn now to the BRST
quantization of the these models. Following the BFV prescription for the nilpotent
BRST charge (see \cite{BF1}) we find the following expression   for the
model \r{202}  \be &&Q=Q_0+\ca(\pi+H_0)+\bapet\pi_v
\e{205}
where $\pi_v$ is the conjugate momentum to $v$, $\ca$  a fermionic ghost
variable and $\bapet$ conjugate momentum to the corresponding 
fermionic antighost $\baca$.
Their fundamental nonzero (graded) commutators are
\be
&&[\ca, \pet]=1,\;\;\;[\baca,
\bapet]=1\;\;\;
[t, \pi]=i,\;\;\;[v, \pi_v]=i
\e{206}
$Q_0$ and $H_0$ are the BFV-BRST charge and BFV Hamiltonian for the model \r{201}.
$Q_0$ and $H_0$ are independent of $\pi$, $v$, $\pi_v$, $\ca$, $\pet$,
$\baca$ and $\bapet$. Since $Q^2=0$ they satisfy (cf.\cite{BLy})
\be
&&Q_0^2=0,\;\;\;[Q_0, \pi+H_0]=0
\e{207}
For instance, in the irreducible case their forms are
\be
&&Q_0=\ca^\al\Phi_\al+\bapet^\al\pi_\al+ \mbox{terms depending on } \pet_\al\nn\\
&&H_0=H+\mbox{terms depending on }\ca^\al\mbox{ and }\pet_\al
\e{208}
where we have introduced the ghosts and antighosts $\ca^\al$,
$\baca_\al$ and their conjugate momenta $\pet_\al$, $\bapet^\al$. $\pi_\al$
are the conjugate momenta to the Lagrange multipliers $v^\al$. All variables are
hermitian and their fundamental nonzero (graded) commutators are (we assume here for
simplicity that $\ca^\al$,
$\baca_\al$, $\pet_\al$, $\bapet^\al$ are fermionic and $\pi_\al$, $v^\al$,
$\Phi_\al$ bosonic) 
\be &&[q^i, p_j]=i\del^i_j,\;\;\;[v^\al,
\pi_\beta]=i\del_\beta^\al,\;\;\; [\ca^\al,
\pet_\beta]=\del^\al_\beta,\;\;\;[\baca_\al, \bapet^\beta]=\del_\al^\beta
\e{209}
We  have of course the usual ghost grading such
that $\ca,\;\bapet,\;\ca^\al,\;\bapet^\al, \; Q, \;Q_0$ have ghost number one,
$\baca,\;\pet, \;\baca_\al,\;\pet_\al$ have ghost number minus one, while
$(q^i,\;p_j)\;(v,\;\pi_v)\;(t,\;\pi)\;(v^\al,\;\pi_\beta)$ and $H_0$ have ghost
number zero. The last conditions imply that the ghost number operator 
\be
&&N=\ca^\al\pet_\al-\baca_\al\bapet^\al+\ca\pet-\baca\bapet
\e{210}
is conserved, \ie $[N, \pi+H_0]=[N, H_0]=0$.

\setcounter{equation}{0}
\section{Time evolution as BRST singlets of the extended formulation}
It is now straight-forward to apply the method of ref.\cite{Solv} to
the above BRST theory of the extended formulation \r{202} of the general gauge theory
\r{201}. This implies that the BRST singlet states may be chosen to have  the form
\r{1}. For the irreducible case \r{208} we have the two expressions
\be &&|s\hb_1=e^{[Q, \psi_1]}|\phi\hb_1,\;\;\;
 |s\hb_2=e^{[Q, \psi_2]}|\phi\hb_2
\e{301}
where $|\phi\hb_{1,2}$ satisfy
\be
&&\chi^\al|\phi\hb_1=\ca^\al|\phi\hb_1=\baca_\al|\phi\hb_1=\pi_\al|\phi\hb_1=0
\e{302}
\be
&&\chi(t)|\phi\hb_1=\ca|\phi\hb_1=\baca|\phi\hb_1=\pi_v|\phi\hb_1=0
\e{303}
\be
&&\Lambda^\al(v^\beta)|\phi\hb_2=\bapet^\al|\phi\hb_2=\pet_\al|\phi\hb_2=[Q,
\pet_\al]|\phi\hb_2=0 
\e{304}
\be
&&\Lambda(v)|\phi\hb_2=\bapet|\phi\hb_2=\pet|\phi\hb_2=(\pi+H_0)|\phi\hb_2=0 
\e{305}
where the hermitian operators $\chi^\al$, $\chi(t)$, $\Lambda^\al(v^\beta)$, and
$\Lambda(v)$ are gauge fixing operators to $[Q,\pet_\al]$, $\pi+H_0$, $\pi_\al$, and
$\pi_v$ respectively. $\chi(t)$, $\Lambda^\al(v^\beta)$, and
$\Lambda(v)$ should therefore be such that the first conditions in
\r{303}-\r{305} fix $t$, $v^\al$, and $v$ uniquely. In particular they must satisfy
$\partial_v\Lambda\neq0$, $\partial_t\chi\neq0$ and
$\det\partial_\al\Lambda^\beta\neq0$ where $\partial_\al\equiv\partial/\partial
v^\al$. (The  most natural choice is of course $\chi(t)=t$,
$\Lambda^\al(v^\beta)=v^\al$, and $\Lambda(v)=v$.) The gauge fixing fermions
$\psi_{1,2}$ in \r{301} may then always be chosen to be \cite{Solv} 
\be
&&\psi_1=\psi_{01}+\pet\Lambda(v),\;\;\;\psi_2=\psi_{02}+\baca\chi(t) 
\e{306}
 where
\be
&&\psi_{01}=\pet_\al\Lambda^\al(v^\beta),\;\;\;\psi_{02}=\baca_\al\chi^\al
\e{307}
are the natural gauge fixing fermions for $Q_0$. 
For this choice  we find
\be
&&[Q,\psi_1]=[Q_0,
\psi_{01}]+\ca[\pi+H_0,\psi_{01}]+\Lambda(v)(\pi+H_0)-i\Lambda'(v)\bapet\pet\nn\\
&&[Q,\psi_2]=[Q_0,
\psi_{02}]+\ca[\pi+H_0,\psi_{02}]+\pi_v\chi(t)-i\dot{\chi}(t)\baca\ca
\e{308}
where $\Lambda'(v)\equiv\partial_v\Lambda(v)$ and
$\dot{\chi}(t)\equiv\partial_t\chi(t)$. The resulting BRST singlets \r{301} do then
represent the general time dependent BRST singlets of the original BRST theory of
\r{201}. In particular one may notice that $|\phi\hb_2$ always satisfy the
Schr\"odinger equation due to the last condition in \r{305}. In the general reducible
or irreducible case \r{302}, \r{304} and \r{307} are replaced by the more general
expressions given in ref.\cite{Solv}. (The last condition in \r{304} may be replaced
by $[Q_0,
\pet_\al]|\phi\hb_2=0$ since
$H_0$ only depends on $\ca^\al$ in terms like $t_\al^\beta\ca^\al\pet_\beta$. Such
reductions are always valid.)

The time evolution of $|s\hb_1$ and $|s\hb_2$ in \r{301} are determined by the
following subset of the  conditions in \r{302}-\r{305}
\be
&&\chi(t)|\phi\hb_1=\Lambda(v)|\phi\hb_2=0
\e{309}
\be
&&\pi_v|\phi\hb_1=(\pi+H_0)|\phi\hb_2=0
\e{310}
The conditions \r{309} in \r{301} imply
\be
&&\chi(t-i\Lambda(v))|s\hb_1=\Lambda(v-i\chi(t))|s\hb_2=0
\e{311}
which relate $t$ and $v$. (Notice that $\chi(t)$ and $\Lambda(v)$ should
be monotonic functions.) The conditions \r{310} in \r{301} yield then generalized
Schr\"odinger equations for $|s\hb_1$ and $|s\hb_2$ when combined with \r{311}. They
are in general rather involved and therefore difficult to express in terms of the
original theory. However, if $\chi(t)$ and $\Lambda(v)$ are chosen to be linear
functions in their arguments, \ie \be
&&\Lambda(v)=\al(v-v_0),\;\;\;\chi(t)=\beta(t-t_0)
\e{312}
where $\al,\beta,v_0,t_0$ are real constants satisfying $\al\neq0,\beta\neq0$, and 
if $\psi_{01}$, $\psi_{02}$ are chosen to be conserved,
\be
&&[\psi_{01}, \pi+H_0]=0,\;\;\;[\psi_{02}, \pi+H_0]=0,
\e{313}
then it is straight-forward to show that \r{310} implies
\be
&&(\pi-i(\Lambda')^{-1}\pi_v+H_0)|s\hb_1=0\nn\\
&&(\pi+i\pi_v\dot{\chi}+H_0)|s\hb_2=0
\e{314}
which when combined with \r{311} reduce to the ordinary  Schr\"odinger
equations. Notice that \r{313} implies
\be
&&[\pi+H_0, [Q_0, \psi_{01(02)}]]=0
\e{3141}
which is the natural generalization of \r{4}. The time evolution seems to be
effectively that of \r{314} even if only \r{312} and \r{3141} are required, since the
modifications in \r{314} will only affect the ghost part of $|s\hb$ in such a way
that they  will not contribute to the inner products. The condition \r{3141} 
also allows  $e^{[Q_0, \psi_0]}$ to be factored out from $e^{[Q, \psi]}$. 
Since the gauge fixing conditions must satisfy Dirac's consistency conditions
the original gauge fixing fermions should be conserved in some weak sense. Whether or
not \r{3141} is the weakest possible conditions remains to determine.

The above reduction to Schr\"odinger equations may be made clearer if we turn to the
wave function representations of $|s\hb_{1,2}$. Inserting \r{308} into \r{301} using
\r{313} we have formally \be
&&\Psi_1(t,v,\pet,\bapet,\omega)=\vb t,v,\pet,\bapet,\omega
|s\hb_1=\nn\\&&=e^{-i\partial_v\Lambda(v)\bapet\pet}
\del(\chi-i\partial_t\chi\Lambda)e^{\Lambda(v)(-i\partial_t+H_0)}e^{[Q_0,
\psi_{01}]}\phi_1(\omega)\nn\\
&&\Psi_2(t,v,\ca,\baca,\omega)=\vb t,v,\ca,\baca,\omega
|s\hb_2=\nn\\&&=e^{-i\partial_t\chi(t)\baca\ca}\del(\Lambda-
i\partial_v\Lambda\chi)e^{[Q_0,
\psi_{02}]}\phi_2(t,\omega) 
\e{315}
 where $\omega$ is a collective coordinate for $q^i,
v^\al,\ca^\al,\baca_\al$ which are the variables of the original BRST theory. In
\r{315} we have used the fact that the wave function representation of $|\phi\hb_{1,2}$
may be written as \be
&&\vb t,v,\pet,\bapet,\omega
|\phi\hb_1=\del(\chi(t))\phi_1(\omega),
\;\;\;\vb t,v,\ca,\baca,\omega
|\phi\hb_2=\del(\Lambda(v))\phi_2(t,\omega)
\e{316} 
where $\phi_1(\omega)$ and $\phi_2(t,\omega)$ satisfy the 
conditions \r{302} and \r{304} respectively.  Eq.\r{316} are explicit solutions of
\r{303} and \r{305} except for the last condition in \r{305} which
requires $\phi_2(t,\omega)$ to satisfy the Schr\"odinger equation
$i\partial_t\phi_2=H_0\phi_2$.

For the natural gauge choice
\r{312} the
wave functions in \r{315} become
\be
&&\Psi_1(t,v,\pet,\bapet,\omega)=e^{-i\al\bapet\pet}
\del(t-t_0-i\al(v-v_0))\,\Phi_1(t,\omega)
\nn\\
&&\Psi_2(t,v,\ca,\baca,\omega)=e^{-i\beta\baca\ca}\del(\beta(t-t_0)+i(v-v_0))
\,\Phi_2(t,\omega)
 \e{317}
where
both $\Phi_1(t,\omega)$ and $\Phi_2(t,\omega)$ satisfy the Schr\"odinger equation
(notice that $e^{[Q_0,
\psi_{01}]}$ and $e^{[Q_0,
\psi_{02}]}$ are conserved)
\be
&&i\partial_t\,\Phi_{1,2}(t,\omega)=H_0\,\Phi_{1,2}(t,\omega)
\e{318}
$\Phi_{1,2}(t,\omega)$ may be identified with the time dependent BRST singlets of the
original BRST theory.

So far the expressions \r{317} for the singlets are only formal. One should  notice
that $q^i,t,v,\ca,\baca,\pet,\bapet,\omega$ in the wave functions \r{315} and \r{317}
are the eigenvalues of the corresponding hermitian operators. The question that
remains to answer is which eigenvalues are real and which are imaginary. 
In  \cite{Proper}
we gave  a general rule  which says that  the existence of solutions of
the form \r{309} requires  bosonic unphysical (gauge) variables and corresponding
Lagrange multipliers to be  quantized in an opposite manner.  This implies that in
\r{315} and \r{317} $t$ and $v$ must be quantized with real and  imaginary eigenvalues
respectively, or vice versa. This follows also since the argument of the delta
functions in \r{315} and \r{317} must be real. The most natural choice is to let $t$
have real eigenvalues and to let  $v$ be quantized with imaginary eigenvalues $iu$,
$u$ real. $v$ in \eg \r{317} must then be replaced by $iu$ after which \r{317} become 
true solutions if $v_0=0$. By means of the properties
 \be &&v|iu\hb=iu|iu\hb,\;\;\;\vb
-iu|=(|iu\hb)^{\dag}\nn\\ &&\vb iu'|iu\hb=\del(u'-u),\;\;\;\int du|-iu\hb\vb -iu|=\int
du|iu\hb\vb iu|=\bett 
\e{319}
of indefinite metric states, the solutions \r{317} become 
\be
&&\Psi_1(t,iu,\pet,\bapet,\omega)=e^{-i\al\bapet\pet}
\del(t-t_0+\al u)\,\Phi_1(t,\omega)
\nn\\
&&\Psi_2(t,iu,\ca,\baca,\omega)=e^{-i\beta\baca\ca}\del(\beta(t-t_0)- u)
\,\Phi_2(t,\omega)
 \e{320}
and we find 
\be
&&_1\vb s|s\hb_1=\int d\omega dt du d\pet d\bapet
\,\Psi_1^*( t,-iu,\pet,\bapet,\omega^*)\Psi_1( t,iu,\pet,\bapet,\omega)=\nn \\&&= \int
d\omega dt du d\pet d\bapet e^{-2i\al\pet\bapet}\del(t-t_0-\al
u)\del(t-t_0+\al u)\,\Phi_1^*(t, \omega^*)\,\Phi_1(t, \omega)=\nn \\&&=\int d\omega
\,\Phi_1^*(t_0, \omega^*)\,\Phi_1(t_0, \omega)>0
\e{321}
\be
&&_2\vb s|s\hb_2=\int d\omega dt du d\ca d\baca
\,\Psi_2^*( t,-iu,\ca,\baca,\omega)\Psi_2( t,iu,\ca,\baca,\omega)=\nn \\&&= \int
d\omega dt du d\ca d\baca e^{-2i\beta\ca\baca}\del(\beta(t-t_0)+ u)\del(\beta(t-t_0)-
u)\,\Phi_2^*(t, \omega^*)\,\Phi_2(t, \omega)=\nn \\&&=\int d\omega
\,\Phi_2^*(t_0, \omega^*)\,\Phi_2(t_0, \omega)>0 \e{322}
This is the expected result since the
choice $\chi(t)=\beta(t-t_0)$ in \r{303} and in $\psi_2$ in \r{306} should be
interpreted as a gauge choice in which we fix $t$ to be $t_0$. The right-hand sides
can of course be further reduced so that all imaginary bosonic coordinates in
$\omega$ are eliminated and we end up with a regular theory with square integrable
wave functions $\Phi$.

The two different gauge fixing $\chi(t)=\beta t$ and $\chi(t)=\beta (t-t_0)$  should be
connected by a gauge transformation
 \be
&&|s, t_0\hb=|s, 0\hb +Q|\chi\hb
\e{323}
where $|\chi\hb$ is restricted by the condition that both $|s, t_0\hb$ and $|s, 0\hb$
are inner product states. In fact, we have
\be
&&|s,t_0\hb=e^{-i(\pi+H_0)t_0}|s,0\hb
\e{3231}
which agrees with \r{323} since $\pi+H_0=[Q,\pet]$. Thus, 
\be
&&\vb s, t_0|s, t_0\hb=\vb s, 0|s, 0\hb 
\e{324}
From \r{321} and \r{322} this requires in turn the time evolution of the original
BRST invariant theory to be unitary. Thus, the different gauge choices are connected by
a unitary gauge transformation exactly as far as the Schr\"odinger equation \r{318}
allows us to connect all time instances in a unitary fashion.  
 Another consequence of \r{323} and \r{3231} is that
\be
&&\vb s, t_0|s, 0\hb=\vb s, t_0|s, t_0\hb=\vb s, 0|s, 0\hb
\e{325}
An explicit calculation using \r{321}  yields also
\be
&&\vb s, t_0|s, 0\hb=\int d\omega\, \Phi^*(t_0/2, \omega^*)\,\Phi(t_0/2, \omega)
\e{326}
in agreement with \r{325} since \r{324} is valid for arbitrary $t_0$.

From the general quantization rule  \cite{Proper} we could also  choose time $t$ to
have imaginary eigenvalues and $v$ real ones.  However, the requirement that the
argument of the delta functions in \r{320} must be possible to be chosen real requires
now $t_0=0$. ($t_0$ cannot be chosen imaginary since $\chi(t)$ must be hermitian.)  
The corresponding gauge fixing is then effected by the hermitian choice $\chi(t)=\beta
t$ and $\Lambda(v)=\al (v-v_0)$.  The solutions in \r{320} become here \be
&&\Psi_1(iu,v,\pet,\bapet,\omega)=e^{-i\al\bapet\pet}
\del(u-\al(v-v_0))\,\Phi_1(iu, \omega)
\nn\\
&&\Psi_2(iu,v,\ca,\baca,\omega)=e^{-i\beta\baca\ca}\del(\beta
u+v-v_0)\, \Phi_2(iu, \omega)
 \e{327}
where $\Phi_{1,2}(iu,\omega)$ satisfy the Schr\"odinger equation with imaginary time
($t=iu$) \be
&&(\partial_u-H_0(iu))\,\Phi_{1,2}(iu, \omega)=0
\e{328}
In this case we find  
  \be &&\vb s, v_0|s, v_0\hb=\int d\omega\,
\Phi^*_{1,2}(0, \omega^*)\,\Phi_{1,2}(0, \omega)>0 
\e{329}
which is independent of $v_0$. Notice that the gauge fixing $\Lambda(v)=\al(v-v_0)$
does not fix  time to an imaginary value which the delta functions in \r{327}
could suggest. Instead, time is fixed to zero in consistency with the  gauge choice
$\chi(t)=\beta t$.

\setcounter{equation}{0}
\section{Path integral formulations}
The representation \r{1} of BRST singlets, \ie $|s\hb=e^{[Q, \psi]}|\phi\hb$,
constitutes a missing link between operator and  path integral quantization of
general gauge theories. The quantization rules in \cite{Proper} 
 may easily be understood when the inner products of \r{1} is written as path
integrals \cite{Path}. The inner products of the extended singlets
\r{301} are of the form 
 \be
&&\vb s'|s\hb=\vb \phi'|e^{\ga[Q,\psi]}|\phi\hb=\int dR' dR\, {\phi'}^*({R'}^*)
\phi(R)\vb R'|e^{\ga[Q,\psi]}|R^*\hb
\e{501}
where $R$ and $R'$ represent all the involved coordinates (some may have imaginary
eigenvalues). $\ga$ is a real constant. Since this expression is independent of the
value of $\ga$ as long as $\ga\neq 0$, we may make the identification $\ga=\tau-\tau'$.
This allows us to interpret the inner product \r{501}  as the transition amplitude  
\be
&&\vb s,\tau'|s, \tau\hb=\int dR' dR\, {\phi'}^*({R'}^*)
\phi(R)\vb R', \tau'|R^*, \tau\hb
\e{502}
where
\be
&&\vb R', \tau'|R^*, \tau\hb=\vb R'|e^{(\tau-\tau')[Q,\psi]}|R^*\hb=
\int DR DP\:
e^{i\int_\tau^{\tau'}\left(iP\dot{R}-\{Q,
\psi\}\right)}
\e{503}
where $P$ are conjugate momenta to $R$.
The last path integral formula is obtained through the time slice formula
(see \cite{Path}). The first equality in \r{503} requires  $[Q, \psi]$ to be
interpreted as a conserved Hamiltonian, which is possible since $[Q, \psi]$ 
has no explicit $\tau$ dependence. Notice that the conditions \r{302}-\r{305} act as
boundary conditions in \r{502}. (Actually \r{502} is independent of the precise choice
of the first conditions in \r{302}-\r{305} since they are gauge fixing conditions.)  
$\{Q, \psi\}$  represents the 'classical' counterpart (Weyl symbol) of \r{308}. As
shown in \cite{Path} $\{Q, \psi\}$ should be real provided  all Lagrange  multipliers
as well as all the ghosts (or antighosts) are chosen to have imaginary eigenvalues.
This is also true here for  imaginary $v$  if $\Lambda(v)=\al v$ in agreement with the
results of the previous section. Thus, \r{503} is a good path integral formula for
$\tau$, $t$ and $\ca$ ($\baca$) real, and $v$, $\baca$ ($\ca$) imaginary. For
imaginary $t$ we must choose an imaginary $\tau$ since most terms in $[Q, \psi]$  have
real eigenvalues when the Lagrange multipliers are chosen to be real as it
must be \cite{Path,Proper}. Here  the exceptions are \eg the terms
$\al(v-v_0)(\pi+H_0)$ in $[Q, \psi_1]$ and $\pi_v\chi(t)$ in $[Q, \psi_2]$.
Consistency requires therefore real $v$ with range of a half-axis (cf.\cite{Proper}).
The explicit time dependence in the original BRST invariant theory should
 probably  also be invariant under $t\rightarrow -t$.

All the
above path integrals may be reduced to the appropriate
path integrals of the original BRST invariant theory (see \cite{Path})
\be
&&\vb s, t'|s, t\hb=\int d\omega' d\omega\, {\phi'}^*({\omega'}^*)
\phi(\omega)\vb \omega', t'|\omega^*, t \hb
\e{504}
where $\phi(\omega)$ only satisfies \r{302} or \r{304} and where
\be
&&\vb \omega', t'|\omega^*, t\hb
=\int D\omega D\pi_\omega\:
e^{i\int_t^{t'}\left(i\pi_\omega\dot{\omega}-H_0-\{Q_0,
\psi_0\}\right)}
\e{505}
where $\omega$ denotes all coordinates of the original BRST invariant theory and
$\pi_\omega$ their conjugate momenta. This reduction may be performed by an integration
over $t,\;\pi,\;v,\;\pi_v,\;\ca,\;\pet$, $\baca,\;\bapet$ in \r{503} for all the above
cases provided $\Lambda(v)$ and $\chi(t)$ have the linear form \r{312}. In particular
one must choose $\Lambda(v)=\al v$ or $\chi(t)=\beta t$ when $v$ or $t$ are imaginary
in agreement with the results of the previous section.

\setcounter{equation}{0}
\section{Application to the regular case} The above results are also valid when the
original theory is an ordinary regular one. Consider a regular theory described by the
phase space Lagrangian \be &&L_0(t)=p_i\dq^i-H \e{401} corresponding to an
unconstrained Lagrangian. This is obviously a special case of \r{201}. This
regular theory may be described by an extended singular phase space Lagrangian of the
form (the corresponding reparametrization invariant theory) 
\be
&&L(\tau)=p_i\dq^i+\pi\dt-v(\pi+H)
\e{402}
A BRST quantization of this singular model leads to the BFV-BRST charge
operator
\be
&&Q=\ca(\pi+H)+\bapet\pi_v
\e{403} 
(All variables are hermitian.) The BRST singlets are
\be &&|s\hb_l=e^{[Q, \psi_l]}|\phi\hb_l,\;\;\;l=1,2
\e{404}
where $|\phi\hb_{1,2}$ satisfy \r{303} and \r{305}, and where the most natural choice
for the gauge fixing fermions $\psi_{1,2}$ is (cf. \r{306})
\be &&\psi_1=\pet\Lambda(v),\;\;\;\psi_2=\baca\chi(t)
\e{405}
Eq.\r{308} reduces here to
\be
&&[Q,\psi_1]=\Lambda(v)(\pi+H_0)-i\Lambda'(v)\bapet\pet\nn\\
&&[Q,\psi_2]=\pi_v\chi(t)-i\dot{\chi}(t)\baca\ca
\e{406}
Conditions \r{303} and \r{305} imply now \r{311} which relates $t$ and $v$,  the
generalized Schr\"odinger equations
\be
&&(\pi-i(\Lambda')^{-1}\pi_v+H+(\Lambda')^{-1}\Lambda''\bapet\pet)|s\hb_1=0\nn\\
&&(\pi+i\pi_v\dot{\chi}+H+\ddot{\chi}\baca\ca)|s\hb_2=0
\e{407}
and the ghost conditions 
\be
&&(\ca-i\Lambda'\bapet)|s\hb_1=(\baca+i\Lambda'\pet)|s\hb_1=0\nn\\
&&(\pet-i\dot{\chi}\baca)|s\hb_2=(\bapet+i\dot{\chi}\ca)|s\hb_2=0
\e{408}
For the linear choice \r{312} the equations for $|s\hb_1$ reduces to
\be
&&(t+i\al
v)|s\hb_1=(\ca-i\al\bapet)|s\hb_1=(\baca+i\al\pet)|s\hb_1=(H+\pi+
\frac{i}{\al}\pi_v)|s\hb_1=0
\e{409}
and for $|s\hb_2$  we have  the
same equations for $\beta=-1/\al$. Thus, for this choice we may set 
$|s\hb\equiv|s\hb_1=|s\hb_2$. The
Schr\"odinger or wave function representation of $|s\hb$ is
\be
&&\Psi_s(q,t,v,\ca,\baca)\equiv\vb q,t,v,\ca,\baca|s\hb
\e{410}
where $q$ represents the physical coordinates $q^i$ in \r{401}. The solutions of the
equations \r{409} (\ie \r{404}) may then be written as 
\be
&&\Psi_s(q,t,v,\ca,\baca)=e^{-i\ca\baca/\al}\del(\frac{t}{\al}+i v)\Phi(q,t)
\e{411}
where $\Phi(q,t)$ satisfies the Schr\"odinger equation
\be
&&(i\partial_t-H_S(t))\Phi(q,t)=0
\e{412}
where $H_S(t)$ is the Schr\"odinger representation of the the Hamiltonian operator
$H(t)$.  So far all equations are formal. To get the true equations we must
consider the appropriate quantization rules. This proceeds exactly along the lines of
 section 3.

\setcounter{equation}{0}
\section{Conclusions}
In section 2 we have shown that the BFV-BRST charge and BFV Hamiltonian for a general
gauge theory, in which the constraints and Hamiltonian also may depend explicitly on
time, are determined by the standard prescription of the BFV-BRST charge for the
corresponding reparametrization invariant gauge theory which then has no explicit
'time' dependence. We have then investigated to what extent the extended BRST singlets
may be interpreted as the time dependent BRST singlets of the original gauge theory.
We have then found that this is possible at least for the sector of  linear gauge
choices for time and a corresponding Lagrange multiplier, and provided the gauge fixing
fermions of the original theory is conserved in some weak sense.  For these linear
gauges the path integrals for the transition amplitudes of the extended models do
also formally reduce to the appropriate transition amplitudes of the original theory. 
Any gauge choice for time within the linear sector yields equivalent results since
there are no topological obstructions in this sector.

The main message of the present communication is, thus,  that 
the time evolution in BRST quantization for general gauge theories is determined by
the BRST quantization of the corresponding reparametrization invariant theory
provided time is fixed  by a linear gauge choice.

\end{document}